\documentclass[12pt, letter]{article}
\usepackage{amsmath,amssymb,amsthm,appendix,booktabs,etoolbox,fullpage,graphicx,mathptmx,multirow,lscape,pgfplots,relsize,sectsty,setspace,url,verbatim,mathtools}
\usepackage{newtxtext,newtxmath}
\usepackage{tikz}
\usepackage{wrapfig}
\usepackage{rotating}
\usepackage{floatflt}
\usepackage{float}
\usepackage{capt-of}
\usepackage[flushmargin]{footmisc}
\usepackage{afterpage}
\usepackage{tabulary}
\usepackage[shortlabels]{enumitem}
\usepackage{xcolor}
\usepackage{subcaption}
\usepackage{caption}
\usepackage{scalefnt}
\usepackage{physics}
\usepackage{titlesec}
\usepackage{textcomp}
\usepackage[normalem]{ulem}
\usepackage[linesnumbered,vlined,ruled, commentsnumbered]{algorithm2e}
\usepackage[margin=1in, total={7in, 9.25in}, top=0.75in]{geometry}
\usepackage{natbib}
\bibpunct[, ]{(}{)}{,}{a}{}{,}%

\emergencystretch=\maxdimen
\sectionfont{\fontsize{12}{14.6}\selectfont}
\subsectionfont{\fontsize{12}{14.6}\selectfont}
\subsubsectionfont{\fontsize{12}{14.6}\selectfont}
\makeatletter
\patchcmd{\@maketitle}{\LARGE}{\bfseries\fontsize{16}{18}\selectfont}{}{}
\makeatother

\title{A Quantum Algorithm Based Heuristic to Hide Sensitive Itemsets}
\date{\vspace{-10ex}}
\titlespacing{\section}{0pt}{\parskip}{-\parskip}
\titlespacing{\subsection}{0pt}{\parskip}{-\parskip}
\titlespacing{\subsubsection}{0pt}{\parskip}{-\parskip}
\AtBeginEnvironment{tabular}{\singlespacing}
\begin{document}

\maketitle
	\begin{center}
	Abhijeet Ghoshal$^{1}$, Yan, Li$^{2}$, Syam Menon$^{3}$, Sumit Sarkar$^{3}$\\
	\begin{footnotesize}
		$^{1}$Gies College of Business, University of Illinois at Urbana Champaign, \\
		$^{2}$College of Engineering, Pennsylvania State University,\\
		$^{3}$Jindal School of Management, University of Texas at Dallas
	\end{footnotesize}

\singlespacing

	\textbf{Abstract}
\end{center}
Quantum devices use qubits to represent information, which allows them to exploit important properties from quantum physics, specifically superposition and entanglement. As a result, quantum computers have the potential to outperform the most advanced classical computers. In recent years, quantum algorithms have shown hints of this promise, and many algorithms have been proposed for the quantum domain. There are two key hurdles to solving difficult real-world problems on quantum computers. The first is on the hardware front -- the number of qubits in the most advanced quantum systems is too small to make the solution of large problems practical. The second involves the algorithms themselves -- as quantum computers use qubits, the algorithms that work there are fundamentally different from those that work on traditional computers. As a result of these constraints, research has focused on developing approaches to solve small versions of problems as proofs of concept -- recognizing that it would be possible to scale these up once quantum devices with enough qubits become available. Our objective in this paper is along the same lines. We present a quantum approach to solve a well-studied problem in the context of data sharing. This heuristic uses the well-known Quantum Approximate Optimization Algorithm (QAOA). We present results on experiments involving small datasets to illustrate how the problem could be solved using quantum algorithms. The results show that the method has potential and provide answers close to optimal. At the same time, we realize there are opportunities for improving the method further.


\doublespacing

\section{Introduction}\label{intro}
\noindent While quantum computing is still in its infancy, its potential has been demonstrated on some well-studied problems. This potential comes from the fact that quantum computers use qubits (quantum bits) to represent information rather than the bits used in classical computers. Qubits are two-state systems based on subatomic particles (typically electrons or photons), and are among the simplest systems that exhibit the fundamental properties of quantum physics. Two of the properties that are particularly relevant to quantum computing are superposition and entanglement. While qubits have two distinct states (representing `0' and `1'), the ability to exist in a superposition of both states simultaneously makes them both different from, and more powerful than, the bits used in classical computers. Entanglement enables the state of a qubit to be changed by changing the state of a perfectly correlated qubit, usually by using an external magnetic field.  \citet{g2019} points out that unlike a traditional computer where ``doubling the number of bits doubles its processing power,'' these properties imply that ``adding extra qubits to a quantum machine produces an exponential increase in its number-crunching ability''. 

The differences between qubits and bits necessitate algorithms developed for quantum computers to be fundamentally different from those designed for classical ones. Researchers have made perceptible progress in developing new quantum algorithms, with the Bernstein-Vazirani algorithm, the Deutsch-Jozsa algorithm, Grover’s algorithm, and Shor’s algorithm~\citep{nc2010} being some of the better known. Currently, the number of qubits in the most advanced systems are still too small to tackle full-scale real-world problems. As a result, the focus of applied research has been on developing prototypes to solve such problems -- prototypes involving approaches that can be scaled up as more qubits become available.
	
One category of algorithms that has shown particular promise employs a combination of classical and quantum computers, and are referred to as Variational Quantum Algorithms (VQAs). Farhi et al. (2014) introduced a VQA called the Quantum Approximate Optimization Algorithm (QAOA) to solve combinatorial problems. QAOA is grounded in adiabatic quantum computing, a form of quantum computing that relies on the well-established adiabatic theorem~\citep{BornFock1928} from quantum mechanics. QAOA uses qubits to encode the decision variables of the optimization problem.  In this paper, we show how QAOA can be used to solve a long standing problem in information systems involving the hiding of sensitive information when sharing data -- the frequent itemset hiding problem. 

The frequent itemset hiding problem arises in the context of retailers sharing transactional data with business partners. Itemsets are sets of items purchased together frequently. Some of the itemsets in the dataset being shared could be sensitive to the data owner (for example, if they were the result of unexpectedly successful sales promotions). While sharing the data could be of benefit to both parties, retailers are likely to do so only if the sensitive itemsets are hidden prior to sharing. The frequent itemset hiding problem involves hiding the sensitive itemsets by altering the fewest number of transactions possible (i.e.,  maximizing the accuracy of the shared database vis-\`{a}-vis the original, unmodified, one). It is known to be NP-hard, and has been well-studied in the literature~\citep[e.g.,][]{vebs2004,msm2005}. 

In this paper, we present an approach to solve the frequent itemset hiding problem using QAOA. \citet{msm2005} formulate it as a generalized assignment problem involving binary variables, a linear objective function, and as many constraints as there are sensitive itemsets. As the structure of QAOA fits quadratic unconstrained binary optimization problems (i.e., unconstrained optimization problems with binary decision variables and an objective function quadratic in the decision variables), QAOA cannot be applied to the frequent itemset hiding problem directly. Consequently, we present a heuristic that adapts our problem to a form that can be solved using QAOA and demonstrate how an existing quantum algorithm can be utilized to solve the frequent itemset hiding problem. By doing so, we contribute to the growing body of research dedicated to methodological improvements for solving real-life problems using quantum algorithms. We also advance research in the domain of data privacy by developing quantum approaches to solve problems relevant to businesses.

We provide an overview of the frequent itemset hiding problem in Section~\ref{sec:fih}. A brief discussion of the current state of quantum computing and QAOA is in Section~\ref{sec:qaoa}. Section~\ref{sec:solution} presents the approach we propose to solve the frequent itemset hiding problem. The results of experiments on small datasets that demonstrate proof of concept are in Section~\ref{sec:exp}. Section~\ref{sec:con} concludes with a discussion on directions for future research.

\section{The Frequent Itemset Hiding Problem}\label{sec:fih}
\noindent A large body of research has attempted to solve the frequent itemset hiding problem (on traditional computers) using a variety of approaches. Among the earlier works is that of~\citet{vebs2004}, who proposed heuristic approaches to hide sensitive itemsets. Over time, many other approaches have been proposed to solve the problem such as methods involving genetic algorithms~\citep{lhyw2015}. However, there is no prior research on the application of quantum algorithms to hide sensitive information.~\citet{msm2005} were the first to present an integer programming formulation. As this is the formulation we consider in this paper, a brief description is provided below.

Given a set of items $\mathcal{J}$, an itemset is any subset $j\subseteq\mathcal{J}$. A transaction is also a subset of items, and a database $\mathcal{D}$ is a set of transactions. The support of an itemset $j$ (i.e., the number of transactions containing $j$) is $\mu_j$.  The set of frequent itemsets is $\mathcal{F}$, where an itemset $j$ is defined to be frequent in $\mathcal{D}$ if the support for $j$ (i.e., the number of transactions containing $j$) is at least equal to a predetermined, user-specified minimum mining support threshold, $\mu_{min}$. The set of sensitive frequent itemsets that the data owner wants hidden is $\mathcal{F}^R$; an itemset $j$ is considered hidden if its support is below a user-specified hiding threshold $\mu^j_h$. While this is not required, we assume that $\mu_j^h=\mu_{min}\ \ \forall j\in\mathcal{F}^R$ as done in most prior work. The frequent itemset hiding problem is formulated as $FIH$ below.
\vspace*{-0.2in}
\begin{align}\label{formulation:FIH}
	\left\{\min \sum_{m\in\mathcal{D}} x_m\big|
	\sum_{m\in\mathcal{D}} a_{mj}x_m\geq \mu_j-\mu_j^h+1\ \ \forall j\in\mathcal{F}^R;
	x_m\in\{0,1\}\ \ \forall m\in\mathcal{D}\right\}\tag{FIH}
\end{align}
\vspace*{-0.4in}

Parameter $a_{mj}=1$ if transaction $m$ supports itemset $j$, and 0 otherwise. Variable $x_m$ is 1 if transaction $m$ is identified for sanitization (sanitization refers to the removal of sensitive itemsets from a transaction), and 0 otherwise. The objective function minimizes the number of transactions sanitized (i.e., maximizes accuracy), while the constraints ensure that all sensitive itemsets are hidden. Transactions that do not support any sensitive itemset will not be sanitized, and can be eliminated from $FIH$ beforehand. For notational convenience, we continue to use $\mathcal{D}$ to represent the database after these transactions have been
removed. Once transactions are identified for sanitization, appropriate items can be removed from them, thereby decreasing the support of sensitive itemsets supported by each of the transactions. 

Consider the dataset in Table \ref{tab:DS-SI}(a) and the sensitive itemsets in Table \ref{tab:DS-SI}(b) from~\citet{msm2005}. The specified hiding threshold for all the itemsets is 3, i.e., $\mu_j^h=3\ \ \forall j\in\mathcal{F}^R$. The associated formulation $FIH$ is in Figure \ref{fig:fih-ds}. Constraints ($\text{cn1}$) -- ($\text{cn5}$) ensure that the five sensitive itemsets in Table~\ref{tab:DS-SI}(b) get hidden. The optimal solution is to sanitize transactions $t_1, t_2, t_4, t_5, t_6$ and $t_9$ (i,e., to set $x_1, x_2, x_4, x_5, x_6$ and $x_9$ to 1); the corresponding objective function value is 6.

\begin{table}[ht]
	\centering
	\begin{tabular}{|c|l|c|l|ll|c|l|l|}
		\cline{1-4} \cline{7-9}
		\multicolumn{1}{|l|}{\textbf{id}} & \textbf{items} & \textbf{id} & \textbf{items} &  &  & \textbf{Itemset} & \textbf{Items} & \textbf{Supported By}       \\ \cline{1-4} \cline{7-9} 
		$t_1$                                & 2,7,8          & $t_6$          & 2,4,7,8        &  &  & $r_1$               & 2,8            & $t_1,t_4,t_5,t_6,t_7$              \\ \cline{1-4} \cline{7-9} 
		$t_2$                                & 0,1,3,4,7,8    & $t_7$          & 0,2,3,4,8      &  &  & $r_2$               & 7,8            & $t_1,t_2,t_4,t_5,t_6,t_7,t_8,t_9,t_{10}$ \\ \cline{1-4} \cline{7-9} 
		$t_3$                          & 0,3,4,8        & $t_8$         & 0,7,8          &  &  & $r_3$          & 0,3,8          & $t_2,t_3,t_4,t_7,t_9$              \\ \cline{1-4} \cline{7-9} 
		$t_4$                          & 0,2,3,4,7,8    & $t_9$         & 0,3,4,7,8      &  &  & $r_4$              & 3,4,8          & $t_2,t_3,t_4,t_7,t_9 $             \\ \cline{1-4} \cline{7-9} 
		$t_5$                          & 1,2,4,7,8      & $t_{10}$      & 3,7,8          &  &  & $r_5$           & 0,3,4,7        & $t_2,t_3,t_4,t_9$                 \\ \cline{1-4} \cline{7-9} 
		\multicolumn{4}{c}{(a) Dataset}&\multicolumn{2}{c}{~~}&\multicolumn{3}{c}{(b) Sensitive itemsets}
	\end{tabular}
	\caption{Example Dataset and Sensitive Itemsets}\label{tab:DS-SI}
	\vspace*{-0.1in}
\end{table}

\singlespacing
\vspace*{-0.2in}
\begin{figure}[ht]
	\[
	\begin{array}{rrrrrrrrrrrrrrrrrrrrrrr}
		\min & x_1&+&x_2&+&x_3&+&x_4&+&x_5&+& x_6&+&x_7&+&x_8&+&x_9&+&x_{10}& & &(\text{obj})\\
		\text{s.t.}&x_1& &   &&   &+&x_4&+&x_5&+ & x_6&+&x_7& &   & &   &&    &\geq&3&(\text{cn1})\\
		&x_1&+&x_2& &   &+&x_4&+&x_5&+ & x_6& &   &+&x_8&+&x_9&+&x_{10}&\geq&6&(\text{cn2})\\
		&   & &x_2&+&x_3&+&x_4& &   & &     &+&x_7& &   &+&x_9& &    &\geq&3&(\text{cn3})\\
		&   & &x_2&+&x_3&+&x_4& &   & &     &+&x_7& &   &+&x_9& &    &\geq&3&(\text{cn4})\\
		&   & &x_2& &   &+&x_4& &   & &     & &   & &   &+&x_9& &    &\geq&1&(\text{cn5})\\
		&x_1&,&x_2&,&x_3&,&x_4&,&x_5&, & x_6&,&x_7&,&x_8&,&x_9&,&x_{10}&\in&\multicolumn{2}{c}{\{0,1\}}\\
	\end{array}
	\]
	\caption{Formulation $FIH$ for Example Dataset}
	\label{fig:fih-ds}
\end{figure}

\doublespacing

\section{Quantum Algorithms}\label{sec:qaoa}
\noindent Variational quantum algorithms are being used to develop prototype algorithms for many real-world problems. Small-scale implementations of algorithms developed by various researchers (e.g., to solve problems related to portfolio optimization and drug discovery) are available as tutorials from IBM~\citep{ibm2023}. In the combinatorial optimization context,~\citet{hgtsbg2021} show how a vehicle routing problem faced by petrochemical companies can be solved using quantum algorithms. This is a fast-growing area with new algorithms being developed at a rapid pace, and QAOA is one of the more popular VQAs with several applications~\citep{l2014}.

\subsection{Definitions and Terms: A Short Primer}
\noindent As noted earlier, a qubit is a two-state system that displays quantum properties. The two orthonormal basis states are typically represented in ``bra-ket'' notation by the vectors $\ket{0}=\bigl[ \begin{smallmatrix}
		1\\0
	\end{smallmatrix} \bigr]$ and $\ket{1}=\bigl[ \begin{smallmatrix}
		0\\1
	\end{smallmatrix} \bigr]$.
Together, the two orthonormal basis states are referred to as the {\em computational basis}. Given the computational basis, the state $\ket{\psi}$ of any qubit can be represented as a linear combination of the basis vectors, i.e., $\ket{\psi}=\alpha_0\ket{0}+\alpha_1\ket{1}$. The Born rule states that measuring qubit  $\ket{\psi}=\alpha_0\ket{0}+\alpha_1\ket{1}$ results in a 0 being observed with probability $\alpha_0^2$ and in a 1 being observed with probability $\alpha_1^2$. Consequently, $\alpha_0^2+\alpha_1^2=1$ for all valid qubit states. 	

All operators on qubits other than the measurement operator are unitary\footnote{A matrix $A$ is called {\itshape unitary} if its conjugate transpose $A^*$ is also its inverse (i.e., $AA^* = A^*A = AA^{-1}= A^{-1}A = I$). The conjugate transpose of a matrix $A$ is obtained by transposing $A$ and applying the complex conjugate of each element of the transpose (where the complex conjugate of $a+ib$ is defined as $a-ib$).} matrices. Measurement operators are Hermitian matrices\footnote{A matrix $A$ is {\itshape Hermitian} if it is equal to its conjugate transpose, i.e., if $A=A^*$.}. Applying an operator on a qubit changes the state of the qubit. The operators relevant to our context are listed below, followed by an illustration of an operator on a qubit. 
\begin{enumerate}[(i), leftmargin=*, noitemsep, topsep=0pt]
	\item A Hadamard gate $H$ is defined to be $\frac{1}{\sqrt{2}}\bigl[ \begin{smallmatrix}
		1 & 1\\1 & -1
	\end{smallmatrix} \bigr]$.
	\item There are two Hamiltonian operators in the context of QAOA, the cost Hamiltonian $H_C$ and the mixer Hamiltonian $H_M$. We discuss these in detail in the next section.
	\item The Pauli-$X$ matrix (gate) is defined to be $\sigma^x=\bigl[\begin{smallmatrix}
		0 & 1\\1 & 0
	\end{smallmatrix} \bigr]$ 
	\item The Pauli-$Z$ matrix (gate) is defined to be $\sigma^z= \bigl[\begin{smallmatrix}
		1 & 0\\0 & -1
	\end{smallmatrix} \bigr]$.
	\item The measurement operator M measures a qubit and outputs a binary value  (0 or 1).
\end{enumerate}
As an example, consider applying a Hadamard gate on a qubit whose initial state is $\ket{\psi}=\ket{0}$. Operators apply on qubits from the left side, so

 \singlespacing
 \vspace*{-0.4in}
\begin{align*}
	H\ket{\psi}=H\ket{0}=\frac{1}{\sqrt{2}} \begin{bmatrix}
		1 & 1\\1 & -1
	\end{bmatrix} \begin{bmatrix}
		1 \\0
	\end{bmatrix}=\frac{1}{\sqrt{2}}\begin{bmatrix}
		1\\1
	\end{bmatrix}=\frac{1}{\sqrt{2}}\left(\begin{bmatrix}
		1\\0
	\end{bmatrix}+ \begin{bmatrix}
		0\\1
	\end{bmatrix}
	\right)=\frac{1}{\sqrt{2}}\left(\ket{0}+\ket{1}\right).
\end{align*}

\doublespacing
Note that $\frac{1}{\sqrt{2}}\left(\ket{0}+\ket{1}\right)$ implies that $\alpha_0 = \alpha_1 = \frac{1}{\sqrt{2}}$. The probabilities of observing a 0 and a 1 are therefore equal (since $\alpha_0^2 = \alpha_1^2 = \frac{1}{2}$), and the qubit is said to be in {\itshape uniform superposition}. The qubit $\ket{\psi}=\ket{0}$ is perfectly aligned with the basis vector $\ket{0}$, which means that the probability of observing 0 when it is measured is 100\%. Applying a Hadamard gate therefore changes its state to one of uniform superposition. Along the same lines, applying a Hadamard gate on $\ket{1}$ changes its state to $\left(\frac{1}{\sqrt{2}}\ket{0}-\frac{1}{\sqrt{2}}\ket{1}\right)$. In this case, $\alpha_0 = \frac{1}{\sqrt{2}}$ and $\alpha_1 = -\frac{1}{\sqrt{2}}$. As $\alpha_0^2 = \alpha_1^2 = \frac{1}{2}$ for this state as well, it is also a uniform superposition state. As these examples illustrate, the state of a qubit is a linear combination of the associated computational basis states.

\subsection{A Brief Discussion of QAOA}
\noindent\citet{fgg2014} proposed QAOA to solve combinatorial optimization problems, illustrating their approach on the well-studied max-cut problem: given a graph $\mathcal{G}\left(\mathcal{V},\mathcal{E}\right)$, partition the vertices in $\mathcal{V}$ into two sets $\mathcal{X}$ and $\mathcal{Y}$ such that the number of edges between $\mathcal{X}$ and $\mathcal{Y}$ is maximized. If we define variable $x_m$ to be 1 if vertex $m$ is placed in partition $\mathcal{X}$ and to be $-1$ if it is placed in partition $\mathcal{Y}$,  the  max-cut problem can be formulated as a quadratic unconstrained binary optimization problem where $\sum_{\{m,n\}\in\mathcal{E}}\frac12\left(1-x_mx_n\right)$ needs to be maximized.
\begin{figure}[htb]
	\begin{center}
		\includegraphics[width=0.9\textwidth]{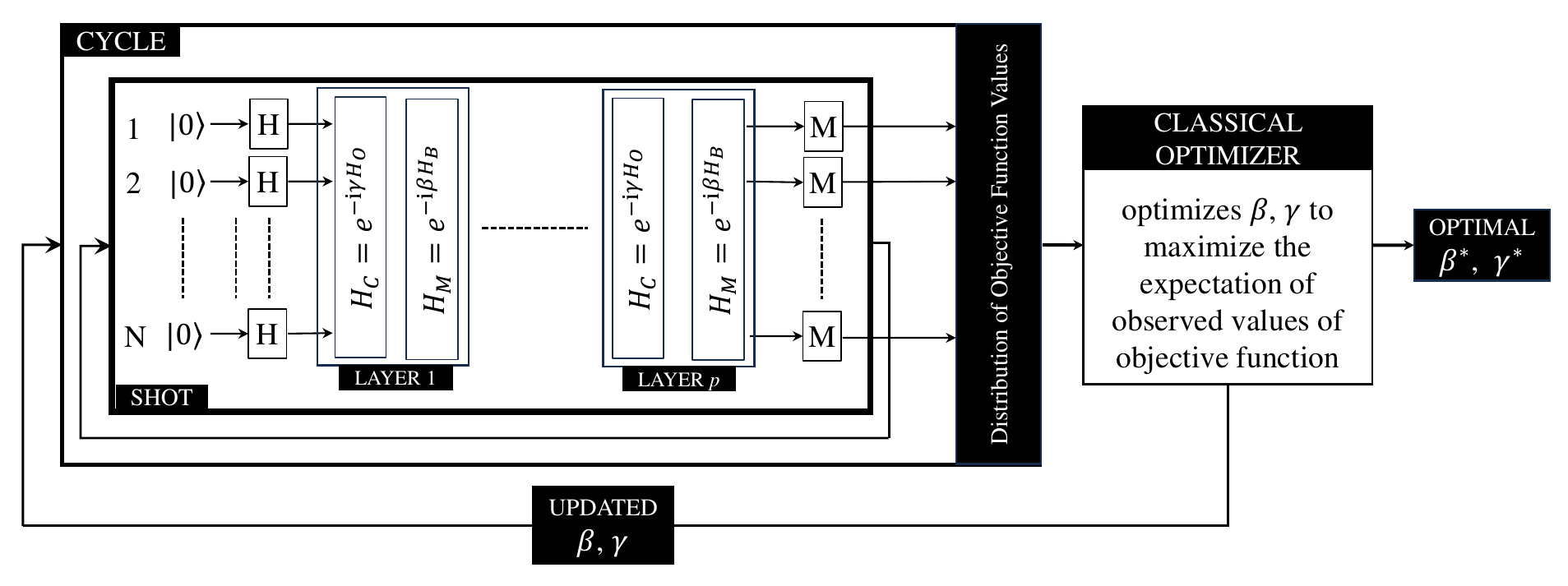}
	\end{center}
	\vspace*{-0.2in}
	\caption{Generic QAOA Flowchart}
	\label{fig:QAOA}
	\vspace*{-0.1in}
\end{figure} 

 A generic flowchart of QAOA is shown in Figure \ref{fig:QAOA}. We discuss it in the context of the max-cut problem considered in~\citet{fgg2014}. The basic idea is to iterate through cycles updating the values of $\beta$ and $\gamma$ until the process converges with optimal values $\beta^*$ and $\gamma^*$. Each cycle involves multiple ``shots'' comprising many steps. 

As qubits correspond to binary decision variables in QAOA, there are $N$ decision variables represented in Figure~\ref{fig:QAOA}. All qubits are initialized to $\ket{0}$ at the beginning of each shot and Hadamard ($H$) gates applied to each, resulting in all qubits being in uniform superposition. At this point, cost and mixer Hamiltonians ($H_C$ and $H_M$ below) are applied on all qubits sequentially in each layer.
\vspace*{-0.2in}
\begin{align}\label{eq:H-QAOA}
	H_C&=e^{-\mathrm i\gamma H_{O}}=e^{-\mathrm i\gamma \sum_{m,n}\frac12\left(I-\sigma^z_m\sigma^z_n\right)}=\displaystyle \prod_{m,n} e^{-\mathrm i\gamma\left(\frac{1}{2}(I-\sigma_m^z\sigma_n^z)\right)}\\
	H_M&=e^{-\mathrm i\beta H_B}=e^{-\mathrm i\beta \sum_{m}\sigma_x}=\displaystyle \prod_{m} e^{-\mathrm i\beta\sigma_m^x}.
\end{align}
\vspace*{-0.5in}

\noindent $\sigma_m^x$ and  $\sigma_m^z$ are Pauli-$X$ and Pauli-$Z$ matrices applied to qubit $m$, while $I$ is the identity matrix. $H_O$ corresponds to the objective function and is set to $\frac12\sum_{m,n}(I-\sigma_m^z\sigma_n^z)$. In the context of the max-cut problem, this corresponds to an application of $(I-\sigma_m^z\sigma_n^z)$ on every pair of qubits $m$ and $n$ such that the corresponding vertices $m$ and $n$ are directly connected by an edge. $H_B$ is set to $\sum_m\sigma_m^x$; this corresponds to the application of $\sigma^x$ on each qubit $m$. The `$\mathrm i$' is $\sqrt{-1}$, the imaginary unit in complex numbers. This completes the first layer. Note that at this stage, the qubits have changed states as a result of the application of the two Hamiltonians. This process of applying the two Hamiltonians $H_C$ and $H_M$ sequentially is repeated for each of the $p$ layers in the model. At this point, all qubits are measured by applying measurement operators M resulting in binary (0 or 1) values for each qubit. Once the qubits are measured, the binary decision values $x_m$ are known for all variables $m$. The corresponding objective function value is $\frac12\sum_{m,n}(1-x_mx_n)$, with $x_m\in\{-1,1\}$ as defined earlier in the context of max-cut problem. The two possible outcomes when a qubit is measured are $x^\prime_m = 0$ and $x^\prime_m=1$, and not -1 and 1. By defining $x_m = (2x_m^\prime -1)$, we get $x_m = -1$ when $x_m^\prime = 0$ and $x_m = 1$ when $x_m^\prime = 1$. The objective function $\frac12\sum_{m,n}(1-x_mx_n)$ is therefore equivalent to $\sum_{m,n}\left(x^\prime_m+x^\prime_n-2x^\prime_mx^\prime_n\right)$ in terms of the observed qubit values. The ``'shot'' referred to earlier comprises the sequence of steps from the initialization of the qubits to their measurement.

Typically, 1,000 shots are run in a cycle to obtain a distribution of  objective function values over the shots in that cycle\footnote{Measurements for different shots would typically be different for each qubit, resulting in different solutions and corresponding objective function values in each shot. As a result, we end up with a distribution of objective function values at the end of each cycle.}. The values of $\beta$ and $\gamma$ remain the same across all the shots in a cycle\footnote{These values are allowed to differ across layers, but we have kept them the same in our discussion for simplicity.}, with both typically initialized to 1 at the beginning of the first cycle. The distribution of objective function values obtained after each cycle is passed to a classical optimizer, which uses non-derivative based algorithms~\citep[e.g.,][]{p1994} to identify new values for $\beta$ and $\gamma$ such that the expectation of the objective function value (based on the distribution) is maximized. These new values of $\beta$ and $\gamma$ are used in the next cycle. This process is repeated until the values of $\beta$ and $\gamma$ converge, with the final $\beta^\ast$ and $\gamma^\ast$ values being used to obtain the final distribution of the objective function value.

While these layers may look similar to layers of neurons in a deep neural network, there is a fundamental difference -- superposition is not feasible on classical computers, and therefore, in classical neural network algorithms. QAOA is a heuristic that uses superposition\footnote{QAOA does not use entanglement.} and other transformations through unitaries to guarantee an optimal solution for quadratic unconstrained optimization problems when the number of layers $p\rightarrow\infty$. As the number of layers have to be limited in practice ($p$ is typically set to 3), the solutions are not guaranteed to be optimal. However, they have been observed to be very close, implying that as more powerful quantum computers become available, the solution of large (real-world) problems using quantum algorithms will become possible.

\section{Ising Formulation-based Heuristic to Solve $FIH$}\label{sec:solution}
\noindent QAOA is suitable for solving problems that can be formulated as Ising models, the standard formulation of which is below~\citep{l2014}. Here, $s_m \in\left\{-1, +1\right\}$, while $J_{mn}$ and $h_m$ are real numbers.

\vspace*{-0.5in}
 \begin{align*}
 	H(s_1,\ldots,s_N)=-\displaystyle\sum_{m<n}J_{mn}s_ms_n-\displaystyle\sum_{m=1}^{N}h_ms_m
 \end{align*}
 \vspace*{-0.5in}

\noindent Note that this maps easily to the structure of the objective function of the quadratic unconstrained binary optimization problem (maximize $\sum_{\{m,n\}\in\mathcal{E}}\frac12\left(1-x_mx_n\right)$ with $x_m\in\{-1,1\}$) that QAOA was designed to solve.

While this formulation does not have constraints, $FIH$ does. In order to apply this approach to the frequent itemset hiding problem, we relax $FIH$ by dualizing the constraints and adding them to the objective function. If $\lambda_j\geq0$ is the multiplier associated with constraint $j$, the Lagrangian dual problem is: 
\vspace*{-0.3in}
\begin{align}
	\mathcal{L}(\lambda): \min & \sum_{m\in\mathcal{D}}x_m+\sum_{j\in\mathcal{F}^R}\lambda_j\left(\left(\mu_j-\mu_j^h+1\right)-\sum_{m\in\mathcal{D}}a_{mj}x_m\right)\\
	&x_m\in\{0,1\}\ \ \forall m\in\mathcal{D}\nonumber
\end{align}
\vspace*{-0.6in}

\noindent The objective function of the dual can be rewritten as
\vspace*{-0.1in}
\begin{align}\label{ks:obj}
	\mathcal{L}(\lambda): &\min\displaystyle \sum_{m\in\mathcal{D}}\left(1-\displaystyle \sum_{j\in\mathcal{F}^R}\lambda_ja_{mj}\right)x_m+\displaystyle \sum_{j\in\mathcal{F}^R}\lambda_j\left(\mu_j-\mu_j^h+1\right).
\end{align}
\vspace*{-0.4in}

\noindent Defining $c_m = \left(1-\sum_{j\in\mathcal{F}^R}\lambda_ja_{mj}\right)$ and $\mu_j^{'}=(\mu_j-\mu_j^h+1)$ for notational convenience, we have
\vspace*{-0.1in}
\begin{align}\label{ks:obj2}
	\mathcal{L}(\lambda): &\min \displaystyle\sum_{m\mathcal{D}}c_mx_m+\displaystyle \sum_{j\in\mathcal{F}^R}\lambda_j\mu_j'.
\end{align}
\vspace*{-0.5in}

\noindent We now square $\mathcal{L}(\lambda)$ to obtain a function compatible with the Ising model. 
\vspace*{-0.2in}
\begin{align}\label{ks:obj-2}
	\mathcal{L}^2(\lambda)=\min \left(\displaystyle\sum_{m\mathcal{D}}c_mx_m+\displaystyle \sum_{j\in\mathcal{F}^R}\lambda_j\mu_j'\right)^2.
\end{align}
\vspace*{-0.4in}

QAOA is designed to handle binary variables that take on values $\{-1,1\}$. As the possible values of the decision variables $x_m$ in $FIH$ are 0 or 1, we define new variables $s_m=2x_m-1$ to obtain variables that are appropriate for QAOA. This mapping ensures that $s_m=-1$ when $x_m=0$ and 1 when $x_m=1$. Substituting $\frac{1+s_m}{2}$ for $x_m$, we get
\vspace*{-0.2in}
\begin{align}
	H_O =\displaystyle\sum_{m\in\mathcal{D}}\left(\frac{1}{4}c_m^2 +\left(\frac{1}{2}\sum_{m\in\mathcal{D}}c_m+\sum_{j\in\mathcal{F}^R}\lambda_j\mu_j^{'}\right)c_m\right)s_m + \sum_{m,n\in\mathcal{D},m\neq n}c_mc_n\frac{s_ms_n}{2}.\tag{IS-FIH}\label{alg:ising}
\end{align}
\vspace*{-0.4in}

One important aspect of this problem is to identify a valid value for $\lambda_j$. We follow~\citet{ghms2020} and set $\lambda_j$ to a constant $\lambda$, where $\left(\lambda=\frac{1}{\sum_{m\in\mathcal{D}}\sum_{k\in\mathcal{F}^R}a_{mk}}\right)$. This value ensures that the combined impact of $\lambda$ across all the variables set to 1 is less than one, which in turn ensures that the solution will be one of the optimal solutions with $\lambda = 0$. In our example, $\sum_{m\in\mathcal{D}}\sum_{k\in\mathcal{F}^R}a_{mk} = 26$, and therefore, $\lambda=\frac{1}{26}$. As $\sum_{j\in\mathcal{F}^R}a_{mj}$ equals the number of sensitive itemsets supported by transaction $m$, this gives priority to transactions that support more sensitive itemsets and leads the dual to choose solutions that are more likely to be part of the solution to $FIH$. Finally, we replace the decision variables $s_m$ with the corresponding Pauli-$Z$ operators $\sigma_m^z$ to obtain the Hamiltonian~\ref{alg:ising-sigma} below, and solve it using QAOA. 

\vspace*{-0.5in}
\begin{align}
	H_O=\displaystyle\sum_{m\in\mathcal{D}}\left(\frac{1}{4}c_m^2 +\left(\frac{1}{2}\sum_{m\in\mathcal{D}}c_m+\sum_{j\in\mathcal{F}^R}\lambda_j\mu_j^{'}\right)c_m\right)\sigma^z_m + \sum_{m,n\in\mathcal{D},m\neq n}c_mc_n\frac{\sigma^z_m\sigma^z_n}{2},\tag{IS-FIH}\label{alg:ising-sigma}
\end{align}
\vspace*{-0.4in}

We select the highest $O$ objective function values from the final distribution of solutions corresponding to $\beta^\ast$ and $\gamma^\ast$, and check the associated solutions for feasibility starting with the highest value. We select the first feasible solution as the final solution to our problem. 


In our illustrative example, the value of $\left(\frac{1}{2}\sum_{m\in\mathcal{D}}c_m+\sum_{j\in\mathcal{F}^R}\lambda_j\mu_j^{'}\right)$ is 5.12. The values of $c_1-c_{10}$ are: $c_1=0.92$, $c_2=0.84$, $c_3=0.92$, $c_4=0.8$, $c_5=0.92$, $c_6=0.92$, $c_7=0.88$, $c_8=0.96$, $c_9=0.84$, and $c_{10}=0.96$. Therefore, the terms corresponding to the first transaction (i.e., for transaction $m=1$) that contribute to $H_O$ are:
\vspace*{-0.2in}
\begin{align*}
	\left(\frac14 (0.92)^2+5.12\times0.92\right)\sigma^z_1+ 0.92\times0.84\times\sigma^z_1\sigma^z_2+0.92\times0.92\times\sigma^z_1\sigma^z_3+0.92\times0.8\times\sigma^z_1\sigma^z_4+\\
	0.92\times0.92\times\sigma^z_1\sigma^z_5+0.92\times0.92\times\sigma^z_1\sigma^z_6+0.92\times0.88\times\sigma^z_1\sigma^z_7+0.92\times0.96\times\sigma^z_1\sigma^z_8+\\
	0.92\times0.84\times\sigma^z_1\sigma^z_9+0.92\times0.06\times\sigma^z_1\sigma^z_{10}.
\end{align*} 
\vspace*{-0.5in}

We add the terms corresponding to the other qubits along similar lines, and the entire expression becomes the cost Hamiltonian in QAOA. The implementation of $\sigma^z_m\sigma^z_n$ is done as illustrated in the qiskit textbook \citep{ibm2023}.  
\section{Dataset and Experiments}\label{sec:exp}
\noindent Quantum algorithms are typically evaluated using simulators. We have used IBM’s {\em qiskit}\footnote{\url{https://qiskit.org}} package in our experiments. In order to keep the simulation times reasonable especially when the underlying combinatorial problem is NP-Hard, experiments using simulators typically involve a small number of qubits (10 to 20)~\citep[e.g.,][]{jwldw2022,vgsaj2020}. Each qubit usually corresponds to a decision variable. In our case, decision variables correspond to transactions and therefore, each transaction has a corresponding qubit whose value should be determined. As a result, we have used datasets with 10 transactions to illustrate the results of our experiments.

\begin{table}[ht]
	\vspace*{-0.3in}
	\centering
	\begin{tabular}{lcc}
		\hline
		\multicolumn{1}{c}{\textbf{Dataset}} & \textbf{\# of items} & \textbf{Average trn length} \\\hline
		Art-DS                               & 7                    & 4.4                         \\
		BMSPOS1                              & 10                   & 4.0                           \\
		BMSPOS2                              & 10                   & 4.3                         \\
		BMSPOS3                              & 10                   & 3.7                        \\
		Art-BMSPOS1                          & 10                   & 5.0                          \\
		Art-BMSPOS2                          & 10                   & 4.7                         \\
		Art-BMSPOS3                          & 12                   & 4.5                         \\\hline
	\end{tabular}
	\caption{Datasets Used In Computational Experiments}\label{tab:DS}
	\vspace*{-0.2in}
\end{table}
The first dataset -- Art-DS -- is the artificial dataset shown in Table~\ref{tab:DS-SI}(a). We also extract three samples -- BMSPOS1, BMSPOS2, and BMSPOS3 -- from BMSPOS (a real dataset from the repository of the first workshop on Frequent Itemset Mining Implementations (FIMI 2003)). These datasets are marginally modified by adding items to some transactions, to create three additional artificial datasets (Art-BMSPOS1, Art-BMSPOS2, and Art-BMSPOS3). Table~\ref{tab:DS} provides the number of items and the average transaction lengths associated with each of these datasets. We consider five sensitive itemsets in Art-DS, with mining and hiding thresholds set to 3. In the remaining datasets we consider 6 sensitive itemsets and set the mining and hiding thresholds to 2. 

\begin{table}[ht]
	\vspace*{-0.3in}
	\centering
	\begin{tabular}{lcc}
		\hline
		\multicolumn{1}{c}{\textbf{Dataset}}& \textbf{Optimal} & \textbf{Ising}\\
		\hline
		Art-DS      & 6 & 8 \\
		BMSPOS1     & 5 & 7 \\
		BMSPOS2     & 4 & 6 \\
		BMSPOS3     & 4 &  7\\
		Art-BMSPOS1 & 6 & 8 \\
		Art-BMSPOS2 & 6 & 8 \\
		Art-BMSPOS3 & 6 & 8 \\\hline
	\end{tabular}
	\caption{Results of Computational Experiments}
	\label{tab:results}
	\vspace*{-0.2in}
\end{table}
Table~\ref{tab:results} provides the results of these experiments. We use CPLEX version 20.1 to find the optimal solutions for each dataset. The optimal and heuristic objective function values are in columns \textbf{Optimal} and \textbf{Ising} respectively. The results for the heuristic are based on the best settings of $\beta$ and $\gamma$, and $O = 50$. The results from the experiments are encouraging as the heuristic solutions are not far from the optimal. However, there is scope for improvement, and we intend to conduct additional experiments with more layers and other initial parameter values of $\beta$ and $\gamma$. We also plan to explore other initial configurations (different initial states of the qubits) and/or a better choice of Lagrangean multipliers.

\section{Conclusions and Future Research}\label{sec:con}
\noindent In this paper, we demonstrate how the problem of hiding sensitive itemsets in a dataset can be solved using existing quantum algorithms. Specifically, we use QAOA to solve the problem. As it cannot be applied directly to our problem, we propose a heuristic that transforms the problem into an Ising model which can be solved using QAOA.  Our experiments show that the approach has promise. We plan to conduct further experiments on larger datasets (of up to 30 transactions) to see how the heuristic performs as datasets get larger. In addition, as mentioned earlier, we intend to explore different initial configurations, initial parameter values, and additional layers. We also plan to conduct experiments on a real quantum computer with 16 qubits. Current quantum computers have several technological hurdles to cross, such as decoherence (stability of states of qubits), error correction, etc. As these issues are resolved over time, the proposed approach will become more viable.  Finally, although several quantum algorithms exist, none of them can be directly used to solve our problem. Therefore, an algorithm-specific adaptation is needed irrespective of the algorithm used, and such an adaptation is usually not obvious. For our problem, we were able to develop such an adaptation using QAOA. Future research can focus on developing appropriate adaptations for other quantum algorithms.

We have initiated investigation into how the distributed version of the frequent itemset hiding problem can be solved using quantum approaches. With appropriate relaxations and aggregations of constraints, this version of the problem can be adapted for solution via the ring copula n-qubit mixer approach proposed by~\citet{vegp2021} to solve the binary knapsack problem. Most of the ongoing and future experiments for the problem addressed here are also relevant for the distributed version of the problem, and we hope to incorporate what we learn from this work into the extension.

 	\bibliography{IH-QC}

\begin{thebibliography}{15}
\expandafter\ifx\csname natexlab\endcsname\relax\def\natexlab#1{#1}\fi
\expandafter\ifx\csname url\endcsname\relax
  \def\url#1{{\tt #1}}\fi
\expandafter\ifx\csname urlprefix\endcsname\relax\def\urlprefix{URL }\fi
\expandafter\ifx\csname urlstyle\endcsname\relax
  \expandafter\ifx\csname doi\endcsname\relax
  \def\doi#1{doi:\discretionary{}{}{}#1}\fi \else
  \expandafter\ifx\csname doi\endcsname\relax
  \def\doi{doi:\discretionary{}{}{}\begingroup \urlstyle{rm}\Url}\fi \fi

\bibitem[{{Born} and {Fock}(1928)}]{BornFock1928}
{Born}, M., V.~{Fock}. 1928.
\newblock Beweis des adiabatensatzes.
\newblock {\it Zeitschrift fur Physik\/} {\bf 51} 165--180.

\bibitem[{Farhi et~al.(2014)Farhi, Goldstone, and Gutmann}]{fgg2014}
Farhi, Edward, Jeffrey Goldstone, Sam Gutmann. 2014.
\newblock A quantum approximate optimization algorithm.
\newblock {\it arXiv preprint arXiv:1411.4028\/} .

\bibitem[{Ghoshal et~al.(2020)Ghoshal, Hao, Menon, and Sarkar}]{ghms2020}
Ghoshal, Abhijeet, Jing Hao, Syam Menon, Sumit Sarkar. 2020.
\newblock Hiding sensitive information when sharing distributed transactional data.
\newblock {\it Information systems research\/} {\bf 31}(2) 473--490.

\bibitem[{Giles(2019)}]{g2019}
Giles, Martin. 2019.
\newblock Explainer: What is a quantum computer?
\newblock {\it MIT Technology Review\/} .

\bibitem[{Harwood et~al.(2021)Harwood, Gambella, Trenev, Simonetto, Bernal, and Greenberg}]{hgtsbg2021}
Harwood, Stuart, Claudio Gambella, Dimitar Trenev, Andrea Simonetto, David Bernal, Donny Greenberg. 2021.
\newblock Formulating and solving routing problems on quantum computers.
\newblock {\it IEEE Transactions on Quantum Engineering\/} {\bf 2} 1--17.

\bibitem[{IBM(2023)}]{ibm2023}
IBM. 2023.
\newblock Qiskit textbook.
\newblock \urlprefix\url{https://qiskit.org/learn/}.

\bibitem[{Jing et~al.(2022)Jing, Wang, Li, Du, and Wu}]{jwldw2022}
Jing, Hang, Ye~Wang, Yan Li, Liang Du, Ziping Wu. 2022.
\newblock Quantum approximate optimization algorithm-enabled der disturbance analysis of networked microgrids.
\newblock {\it 2022 IEEE Energy Conversion Congress and Exposition (ECCE)\/}. IEEE, 1--5.

\bibitem[{Lin et~al.(2015)Lin, Hong, Yang, and Wang}]{lhyw2015}
Lin, Chun-Wei, Tzung-Pei Hong, Kuo-Tung Yang, Shyue-Liang Wang. 2015.
\newblock The ga-based algorithms for optimizing hiding sensitive itemsets through transaction deletion.
\newblock {\it Applied Intelligence\/} {\bf 42} 210--230.

\bibitem[{Lucas(2014)}]{l2014}
Lucas, Andrew. 2014.
\newblock Ising formulations of many np problems.
\newblock {\it Frontiers in physics\/} {\bf 2} 5.

\bibitem[{Menon et~al.(2005)Menon, Sarkar, and Mukherjee}]{msm2005}
Menon, Syam, Sumit Sarkar, Shibnath Mukherjee. 2005.
\newblock Maximizing accuracy of shared databases when concealing sensitive patterns.
\newblock {\it Information Systems Research\/} {\bf 16}(3) 256--270.

\bibitem[{Nielsen and Chuang(2010)}]{nc2010}
Nielsen, Michael~A, Isaac~L Chuang. 2010.
\newblock {\it Quantum computation and quantum information\/}.
\newblock Cambridge university press.

\bibitem[{Powell(1994)}]{p1994}
Powell, Michael~JD. 1994.
\newblock {\it A direct search optimization method that models the objective and constraint functions by linear interpolation\/}.
\newblock Springer.

\bibitem[{Van~Dam et~al.(2021)Van~Dam, Eldefrawy, Genise, and Parham}]{vegp2021}
Van~Dam, Wim, Karim Eldefrawy, Nicholas Genise, Natalie Parham. 2021.
\newblock Quantum optimization heuristics with an application to knapsack problems.
\newblock {\it 2021 IEEE International Conference on Quantum Computing and Engineering (QCE)\/}. IEEE, 160--170.

\bibitem[{Verykios et~al.(2004)Verykios, Elmagarmid, Bertino, Saygin, and Dasseni}]{vebs2004}
Verykios, Vassilios~S, Ahmed~K Elmagarmid, Elisa Bertino, Y{\"u}cel Saygin, Elena Dasseni. 2004.
\newblock Association rule hiding.
\newblock {\it IEEE Transactions on knowledge and data engineering\/} {\bf 16}(4) 434--447.

\bibitem[{Vikst{\aa}l et~al.(2020)Vikst{\aa}l, Gr{\"o}nkvist, Svensson, Andersson, Johansson, and Ferrini}]{vgsaj2020}
Vikst{\aa}l, Pontus, Mattias Gr{\"o}nkvist, Marika Svensson, Martin Andersson, G{\"o}ran Johansson, Giulia Ferrini. 2020.
\newblock Applying the quantum approximate optimization algorithm to the tail-assignment problem.
\newblock {\it Physical Review Applied\/} {\bf 14}(3) 034009.

\end{thebibliography}
\bibliographystyle{ormsv080}


\end{document}